\documentclass{iaus}
\usepackage{graphicx}

\title[The promise of Gaia] 
{The promise of Gaia \\ and how it will influence stellar ages}

\author[Carla Cacciari ]   
{Carla Cacciari}

\affiliation{INAF, Osservatorio Astronomico di Bologna, \\ 
Via Ranzani 1, 40127 Bologna, Italy \\ 
email: {\tt carla.cacciari@oabo.inaf.it}}

\pubyear{2008}
\volume{258} 
\pagerange{001--010}
\jname{The Ages of Stars}
\editors{E.E. Mamajek, D.R. Soderblom \& R.F.G. Wyse, eds.}
\begin{document}

\maketitle

\begin{abstract} The Gaia space project, planned for launch in 2011,
is one of the ESA cornerstone missions, and will provide astrometric,
photometric and spectroscopic data of very high quality for about one
billion stars brighter than V=20. This will allow to reach an
unprecedented level of information and knowledge on several of the
most fundamental astrophysical issues, such as mapping of the Milky
Way, stellar physics (classification and parameterization), Galactic
kinematics and dynamics, study of the resolved stellar populations in
the Local Group, distance scale and age of the Universe, dark matter
distribution (potential tracers), reference frame (quasars,
astrometry), planet detection, fundamental physics, Solar physics,
Solar system science.\\ I will present a description of the instrument
and its main characteristics, and discuss a few specific science cases
where Gaia data promise to contribute fundamental improvement within
the scope of this Symposium.
\end{abstract}

\firstsection 
\section{Introduction}\label{s:intro}

The idea of measuring the position of stars on the sky systematically
and for a scientific purpose dates back to the 2nd Century BC, when
the Greek astronomer Hipparchus measured about 1000 stars naked
eye. This has been repeated several times during the following
centuries, with steadily increasing power and accuracy, until the most
recent Hipparcos satellite (1989-1993) that measured $\sim$ 120,000
sources with $\sim$ 1 mas accuracy and produced a Catalogue
(\cite[Perryman et al. 1997]{per97}) later revised by
\cite[van Leeuwen (2007)]{fvl07}. 

Gaia represents the natural follow up of the Hipparcos mission, with
huge improvement in terms of: i) measurement accuracy, ii) limiting
magnitude and hence number of observed objects, and iii) the
combination of nearly simultaneous astrometric, photometric and
spectroscopic observations.

Gaia is a cornerstone mission of the ESA Space Program, that will
perform an all-sky survey and produce accurate astrometry and
photometry for about 1.5$\times$10$^9$ objects down to a limiting
magnitude of 20~mag, and additional spectroscopy for objects brighter
than V=16-17. This will allow to obtain a stereoscopic and kinematic
view of the Galaxy, and to address key questions of modern
astrophysics regarding the formation and evolution of the Milky Way.
Such an observational effort has been compared to the mapping of the
human genome for the impact that it will have in Galactic
astrophysics.  In addition, Gaia will provide a fundamental
contribution in a much broader range of scientific areas (see
Sect. \ref{s:science}).

More information on the Gaia mission and its science can be found in
the \cite[{\it Gaia Concept and Technology Study Report}
(2000)]{ctsr}, the Proceedings of the Symposium \cite[``The
3-Dimensional Universe with Gaia'' (2005)]{3D}, and at {\bf
http://www.rssd.esa.int/Gaia}.

In Table \ref{t:compare} Gaia and Hipparcos characteristics and
performances are compared.

\begin{table}
  \begin{center}
  \caption{Comparison of Hipparcos and Gaia characteristics and performance}
  \label{t:compare}
 {\scriptsize
  \begin{tabular}{|l|c|c|} 
  \hline 
  & {\bf Hipparcos } & {\bf Gaia}  \\ 
  \hline
Completeness to ...   & V $\sim$ 9 & V $\sim$ 20 (blue) - 22 (red) \\ \hline
Magnitude limit & V$\sim$12.4     &  $\sim$ 1 mag. fainter than completeness \\ \hline
N. Sources      & $\sim$1.2 10$^5$ & $\sim$1.5 10$^9$  \\
~~~~~ Quasars   & 0          & $\sim$10$^6$ \\
~~~~~ galaxies  & 0          & $\sim$10$^7$ \\ \hline
Astrometric accuracy         & $\sim$ 1 mas & $\le$ 7 $\mu$as at V$\le$10 \\ 
                             &              & 12(red)-25(blue) $\mu$as at V=15 \\
		             &              & 100(red)-300(blue) $\mu$as at V=20  \\ \hline
Photometry                   & 2 bands      & ~~~low-res spectrophotometry, 330-1050 nm ~~~\\ \hline
Spectroscopy (R$\sim$ 11,000)~~~  & none    & 1-10 km s$^{-1}$ at V $\le$ 16(blue)-17(red) \\ \hline
Target selection     & ~~~input catalogue~~~ & real-time onboard selection \\ \hline
  \end{tabular}
  }
 \end{center}
\end{table}

\section{Overview of the Gaia mission}\label{s:overview}

\begin{figure}[h]
\begin{center}
 \includegraphics[width=4.5in]{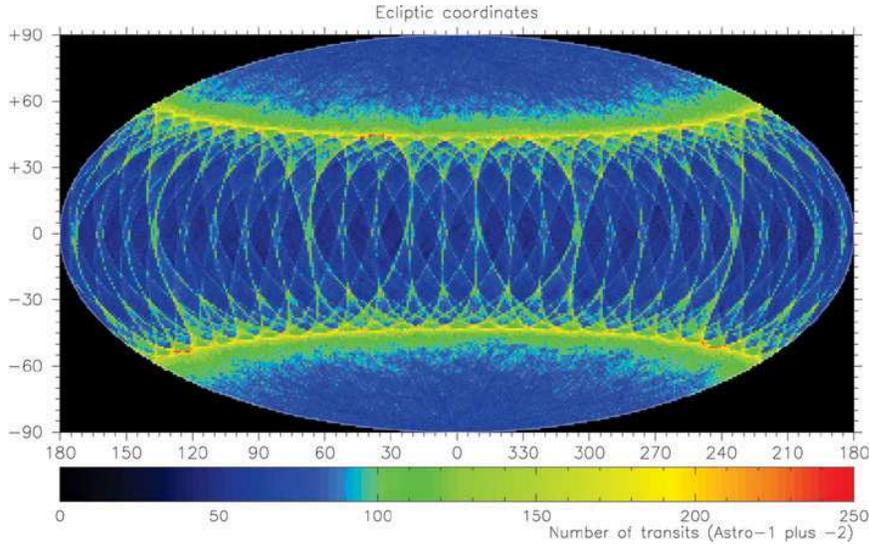}
\caption{
Dependence of the total end-of-mission number of focal plane transits
on position on the sky. Shown is an all-sky equal-area Hammer
projection in ecliptic coordinates.  The maximum number of transits
will occur in a $\sim$ 10-deg wide strip around ecliptic latitudes
+/-- 45 deg.  }
\label{f:scanlaw}
\end{center}
\end{figure}

Gaia is scheduled for launch at the end of 2011 from Kourou by a
Soyuz-Fregat launcher, that will put it in a Lissajous-type
eclipse-free orbit around L2 point of the Sun-Earth system. The design
lifetime is 5 years, with a possible extension of one year. The
satellite will perform a continuous scanning of the sky at a rate of
60 arcsec s$^{-1}$, with a precession period of the spin axis of 70
days. As a result of this scanning law, at the end of the mission the
whole sky will have been observed several times, from a few tens to
more than 200 depending on the position (see Fig. \ref{f:scanlaw}),
the average value being around 80.

\subsection{Measurement Principle}\label{s:measure} 

The mission is designed to perform global (wide field) astrometry as
opposed to local (narrow field) astrometry. In local astrometry, the
star position can only be measured with respect to a neighbouring star
in the same field. Even with an accurate instrument, the propagation
of errors is prohibitive when making a sky survey. The principle of
global astrometry is to link stars with large angular distances in a
network where each star is connected to a large number of other stars
in every direction.

Global astrometry requires the simultaneous observation of two fields
of view in which the star positions are measured and
compared. Therefore, the payload will provide two lines of sight,
obtained with two separate telescopes. Then, like Hipparcos, the two
images will be combined, slightly spaced, on a unique focal plane
assembly. Objects are matched in successive scans, attitude and
calibrations are updated, and object positions are solved and fed back
into the system. The procedure is iterated as more scans are added
(Global Iterative Solution). In this way the system is
self-calibrating by the use of isolated non variable point sources
that will form a sufficiently large body of reference objects for most
calibration purposes, including the definition of the celestial frame.
Extragalactic objects (e.g. QSOs) will be used to attach this to the
International Celestial Reference Frame.

\subsection{Instruments and Data Products}\label{s:inst}

The payload consists of a toroidal structure (optical bench) holding
two primary mirrors whose viewing directions are separated by 106.5
deg (the Basic Angle). These two fields of view get superposed and
combined on the same focal plane, that contains: \begin{itemize}
\item the Sky Mapper: an array of 2$\times$7 CCDs for on-board star detection and 
selection; 
\item  the Astrometric Field (AF): an array of 9$\times$7 CCDs for astrometric 
measurements and integrated white-light photometry;
\item  the Blue (BP) and Red (RP) Photometers:  two columns of 7 CCDs each for 
prism spectrophotometry in the 330-680 nm wavelength range with  dispersion 3-29 nm/px, 
and in the 640-1050 nm wavelength range with  dispersion 7-15 nm/px, respectively; 
\item  the Radial Velocity Spectrometer (RVS): an array of 3$\times$4 CCDs for slitless 
spectroscopy (through grating and afocal field corrector) 
at the Ca II triplet (847-874 nm) with resolution R $\sim$ 11,000. 

\end{itemize}

\begin{figure}[h]
\begin{center}
 \includegraphics[width=4.9in]{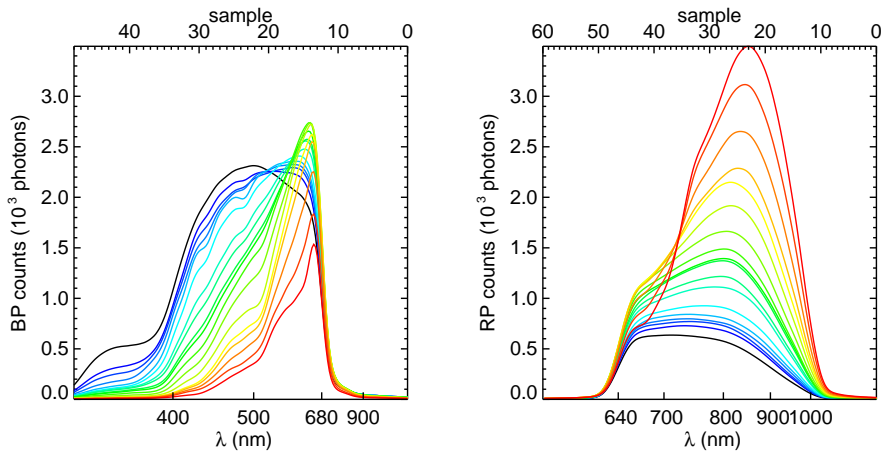}
\caption{Examples of simulated BP and RP prism spectra  
for main sequence stars from O5 to M6, at G=15 and A$_V$=0 
(\cite[Strai$\check{z}$ys et al. 2006]{stra06}).
}
\label{f:spectra}
\end{center}
\end{figure}

Therefore the data produced by Gaia will be of three types:
\begin{itemize}
\item Astrometry (parallaxes, proper motions);
\item Photometry, both integrated (such as the white-light 
$G$-band from the Astrometric Field and the $G_{BP}$ and $G_{RP}$ from 
the blue and red photometers) and low-dispersion  BP and RP spectra.  
We show in Figure \ref{f:spectra} examples of simulated  BP and RP  
spectra as will be observed by Gaia, for main sequence stars from O5 to M6, 
at G=15 and zero reddening (\cite[Strai$\check{z}$ys et al. 2006]{stra06}). 
\item Spectroscopy (radial velocities; rotation, chemistry
for the brighter sources). 
\end{itemize}

\subsection{Performances}\label{s:perf}

\subsubsection{Astrometry}\label{s:perfastrom} 

Astrometric errors are dominated by photon statistics.  The expected
accuracies of parallax measures as a function of V are listed in Table
\ref{t:compare}, and are also reported as a function of M$_V$
(i.e. distance) in Table \ref{t:accur} (left part), along with the
corresponding number of objects that can be observed by Gaia.  We
consider V=7 mag as the bright magnitude limit for astrometric
observations, as saturation sets in at about V=6 mag. \\ Accuracy on
proper motions is about 20\% better than on parallaxes. \\ We note
that comparable astrometric accuracy is (will be) obtained from other
(present) future space and ground-based facilities, but on {\bf
pencil-beam areas} of the sky. \\

\begin{table}
  \begin{center}
  \caption{Accuracy of astrometric and photometric data. 
  Left: parallax accuracy (and number of objects) as a function of M$_V$ and V (and hence distance). 
  Right: accuracy of internally calibrated integrated photometry as a function of $G$ mag.
  }
  \label{t:accur}
  \begin{tabular}{|c|c|c|c|c|c|c|c|c|c|c|c|c|} 
  \hline 
$\sigma(\pi)/\pi$&M$_V$&{\bf -5 }&{\bf 0}&{\bf 5}&{\bf 10}&{\bf 15}&~~~&{\bf ~~$G$~~} &{\bf ~13~}&{\bf ~15~}&{\bf ~17~}&{\bf ~20~} \\ 
 N. of objects     &         &           &         &         &          &         &~~~& mag       &          &          &         &     \\
 \hline
  0.1-0.3\%           & V     &     & 7    & 12  & 15  & 17 &~~~& ~~$\sigma G$~~ & ~0.1~ & ~0.2~ & ~0.8~ & ~2.5~ \\ 
$\sim$ 10$^5$        & D(pc) &     & 250  & 250 & 100 & 25 &~~~& mmag &&&& \\ \hline
 1-3\%              & V     & 7        & 12   & 15  & 17  & 20 &~~~& ~~$\sigma G_{BP}$~~ & ~0.3~ & ~0.8~ & ~6.5~ & ~37.8~ \\ 
$\sim$ 10$^7$        & D(pc) & 2500  & 2500 & 1000 & 250 & 100 &~~~& mmag &&&& \\ \hline
 10-30\%             & V     & 12       & 15   & 17  & 20  &  &~~~& ~~$\sigma G_{RP}$~~ & ~0.3~ & ~0.7~ & ~6.1~ & ~35.2~ \\ 
$\ge$ 10$^8$         & D(pc) & 25000  & 10000 & 2500 & 1000 &   &~~~& mmag &&&& \\ \hline
  \end{tabular}
 \end{center}
\end{table}

\subsubsection{Photometry}\label{s:perfot} 

We show in Table \ref{t:accur} (right part) a summary of the accuracy
that Gaia is expected to achieve for the internally calibrated
integrated photometry in white light ($G$-band) and in the $G_{BP}$
and $G_{RP}$ bands, as a function of $G$ mag.  The quoted values are
predicted end-of-mission (80 transits) mean values and include
Poisson, background and readout noise. A systematic $\sim$ 1 mmag
calibration error should be considered in addition (\cite[Jordi et
al. 2007]{cj042}).  The accuracy per pixel of the BP/RP
spectrophotometric data will be at least one order of magnitude worse,
depending on the shape of the spectral energy distribution.

 
The external (absolute) flux calibration of the Gaia photometric
system will be derived from constraining observations of
spectrofotometric standard stars (SPSS). To this purpose a grid of
SPSS is being built, based on CALSPEC stars and incremented by
additional stars to cover adequately the widest possible spectral type
range.  An observing campaign has been started to ensure accurate and
homogeneous data for a suitably large number of SPSS.  The accuracy of
the absolute calibrated data is expected to be $\sim$ 1\% to a few
percent, depending on the number and spectral type of the SPSS and on
the accuracy of their SEDs, and may vary with wavelength across the
BP/RP spectra.

From the BP/RP spectral energy distributions it will be possible to
estimate astrophysical parameters using pattern recognition techniques
(\cite[Bailer-Jones 2008]{bj08}).  For example, one expects to obtain
(r.m.s. are internal uncertainties at V=15):
\begin{itemize}
\item T$_{eff}$ to 1-5 \% for a wide range of T$_{eff}$; 
\item log$g$ to 0.1-0.4 dex, $<$ 0.1 dex for hot stars (SpT $\le$ A); 
\item $[Fe/H]$ to $<$ 0.2 dex for cool stars (SpT $>$ F) down to [Fe/H]=--2.0 dex;
\item A$_V$ to 0.05-0.1 mag for hot stars;
\end{itemize} thus providing a complete characterisation of stellar
populations.
  
\subsubsection{Spectroscopy}\label{s:perspec} 

The RVS provides the third component of the space velocity of each red
(blue) source down to about 17th (16th) magnitude.

Radial velocities are the main product of the RVS, with typical
accuracies of 1 to 10 km s$^{-1}$ down to the limiting magnitude.  For
brighter sources ($<$ 14 mag) the RVS spectra will provide information
also on rotation and chemistry, and will allow to obtain more detailed
and accurate astrophysical parameters than using the prism BP/RP
spectra alone.

\subsection{Science Data Processing: the DPAC}\label{s:DPAC}  

All aspects of the Gaia mission are charge and responsibility of ESA,
except the processing and analysis of the science data that are
assigned to the European astronomical community. To this purpose the
community has formed the Data Processing and Analysis Consortium
(DPAC), that collects the contribution of nearly 400 scientists from
24 Institutes of ESA member States, and is structured in 9
Coordination Units dealing with all aspects of the data processing.
In addition, a number of Data Processing Centers (DPCs) are dedicated
to the data handling and processing of specific parts of the pipeline,
namely: ESAC (Spain), CNES (France) and the DPCs in Barcelona, Torino,
Toulouse, Cambridge and Geneva.

A final data Catalogue will be produced around 2019-2020, containing
the end-of-mission measurements for the complete sample of objects
down to V=20 mag.  Intermediate catalogues might be released before
the end of the mission, as appropriate. Science alerts data are
released immediately.

No proprietary data rights are implemented.

\section{Science with Gaia}\label{s:science} 

The primary goal of the Gaia mission is to obtain data which allow for
studying the structure, composition, formation and evolution of the
Galaxy.  The detailed knowledge of the Galaxy will provide a firm base
for the analysis of other galaxies for which this level of accuracy
cannot be achieved through direct observations. However, a large
number of objects external to the Galaxy will be reached by the Gaia
instruments, yielding results of no less interest and importance.

\subsection{Science products}\label{s:sciprod}

{\underline {In the Galaxy:}} Gaia will provide a complete census of
all stellar populations down to 20th magnitude.  Based on the Besan{\c
c}on Galaxy model (\cite[Robin et al. 2003, 2004]{rob03}) Gaia is
expected to measure about 9$\times$10$^8$ stars belonging to the Disk,
4.3$\times$10$^8$ Thick Disk stars, 2.1$\times$10$^7$ Spheroid stars
and 1.7$\times$10$^8$ Bulge stars. Binaries, variable stars and rare
stellar types (fast evolutionary phases) will be well sampled, as well
as special objects such as Solar System bodies ($\sim 10^5$),
extra-solar planets ($\sim 2\times10^4$), WDs ($\sim 2\times10^5$),
BDs ($\sim 5\times10^4$).

One billion stars in 5-D (6-D if the radial velocity is available, and
up to 9-D if the astrophysical parameters are know as well) will allow
to derive the spatial and dynamical structure of the Milky Way, its
formation and chemical history (e.g. by detecting evidence of
accretion/merging events), and the star formation history throughout
the Galaxy.  The huge and accurate database will provide a powerful
testbench for stellar structure and evolution models.  The possibility
to obtain clean Colour-Magnitude (and hence HR) diagrams throughout
the Galaxy will lead to accurate mass and luminosity functions, as
well as complete characterisation and dating of all spectral types and
Galactic stellar populations.  The distribution and rate of
microlensing events will allow to map the dark matter
distribution. The cosmic distance scale will get a definitive and
robust definition (zero-point) thanks to the very accurate distance
(i.e. luminosity calibration) of the primary standard candles, RR
Lyraes and Cepheids.

{\underline {Outside the Galaxy:}} the brightest stars in nearby (LG)
galaxies will be observed by Gaia, as well as SNe and burst sources
($\sim 2\times10^4$), distant galaxies ($\sim 10^7$), QSOs ($\sim
10^6$), gravitational lensing events ($\le 10^2$ photometric, a few
$10^2$ astrometric).

{\underline {Fundamental physics and general relativity}} will also
benefit from Gaia observations: as an example, the parameter $\gamma$,
representing the deviation from Newtonian theory of the gravitational
light bending, will be measured to $\sim$ 5$\times$10$^{-7}$ as
compared to the present accuracy of 10$^{-4}$-10$^{-5}$.

\subsection{The impact of Gaia on stellar ages: a few examples}\label{s:age}

\subsubsection{Globular Clusters}\label{s:GC}

Globular clusters (GCs) are among the densest fields in the sky, and
considering that the maximum density that Gaia can handle is $\sim$
0.25 star arcsec$^{-2}$ not all of the 150 GCs known in the Galaxy
(\cite[Harris 1996]{har96}) will be completely observable right to the
centre.  Simulations with King models and concentration parameter
c=0.5 to 2.5 have shown that 30\% of them can be fully observed, most
of the remaining ones can be observed at radial distances r $>$ 1
arcmin, and only 5 will be observable at r $>$ 3 arcmin.  Therefore,
for more than 100 GCs it will be possible to observe from 10$^3$ up to
10$^6$ stars each.

The availability of parallaxes and proper motions, as well as radial
velocities for V$<$ 16-17, allows to assess the membership and hence
to derive clean CM diagrams. At the limiting magnitude V=20,
corresponding to $\ge$ 1 mag below the main-sequence turn-off for the
30 GCs closer than 10 kpc, the accuracy on proper motions is expected
to be $\sim$ 0.08 mas (for red stars) to $\sim$ 0.25 mas (for blue
stars).

For comparison, we note that \cite[King et al. (1998)]{king98}, in
their seminal work on NGC 6397, were able to perform a very good
cleaning of the main-sequence using proper motions of accuracy $\sim$
10 mas obtained from WFPC2-WFC data over a 32 month time baseline. The
most recent achievement by
\cite[Anderson \& King (2006)]{ak06} is $\sim$ 0.5 mas astrometric 
accuracy from ACS/WFC data on well exposed images. This accuracy 
is indeed getting close to Gaia's, but on a field of view of only 
about 200$\times$200 arcsec$^2$. 

Clean CMD are the essential tool to derive the {\em absolute} (and
relative) age of a GC, as we discuss in more detail in the following
section using M3 as an example.

\subsubsection{The age of M3}\label{s:M3}

At a distance of about 10 kpc, the brightest features of M3 CM
diagram, namely the upper RGB and the HB, are at V $\sim$ 12.5 to 16
mag. Therefore these stars will be observed with {\em individual}
accuracy $\sigma(\pi)/\pi \sim$ 7-30\%.  By averaging the results from
1000 such stars distributed according to the RGB luminosity function
(\cite[Ferraro et al. 1997]{ferr97}), the distance to M3 can be known
to about 0.5\% or better.

The most classical clock provided by stellar evolution theory for
dating Population II stars is the luminosity of turn-off stars
M$_V$(TO).  This has been parameterised by \cite[Renzini (1993)]{ren}
as:\\ 

$Logt_9 \propto 0.37M_V(TO)-0.43Y-0.13[Fe/H]$ \\ 

M$_V$(TO) is sensitive to input physics and assumptions that affect
the size and energy production of the radiative core. Comparison of
the various most recent M$_V$(TO) vs. age relations shows that an
intrinsic - and hence systematic - theoretical error in the age
determination may be present.  We refer to the presentation by
B. Chaboyer (this conference) for more details on the
intrinsic/systematic errors of theoretical models.

In addition to this, errors on the observable parameters entering the
M$_V$(TO)-age relation must be considered. By assuming typical values
currently obtained for these errors, the error budget can be
summarised as:\\
\noindent $\bullet$  0.85$\sigma$(V$_{TO}$): error associated to 
the photometric determination of the TO.  Extremely accurate and well
defined main sequences can presently be obtained with instruments such
as the HST and other large ground-based facilities, however the
isochrones are nearly vertical at the TO, and V$_{TO}$ is rather
difficult to define. We assume $\sigma$(V$_{TO}$) $\sim$0.10 mag; \\
\noindent $\bullet$ 0.85$\sigma$(mod):  error associated to the 
distance determination, we assume $\sim$0.10 mag; \\
\noindent $\bullet$ 0.85$\sigma$(A$_V$): error associated to the 
extinction determination, we assume $\sim$0.06 mag; \\
\noindent $\bullet$ 0.99$\sigma$(Y): error associated to the 
Helium abundance determination, we assume  $\ge$0.02 dex; \\
\noindent $\bullet$  0.30$\sigma$[M/H]:  error associated to the 
chemical abundance determination, we assume $\sim$0.10 dex. 

The final accuracy on age determinations is $\sim$13\%, as also
estimated by A. Sarajedini (this conference). In a few particular
cases and with especially accurate data and analysis, the accuracy on
age determination has been claimed to be as low as $\le$10\%
(\cite[Gratton et al. 2003]{gra03}). However, before further
improvement can be achieved systematic errors need to be solved, for
example on chemical abundance determination (by defining the
calibrating Solar mixture and metallicity scale), on helium abundance
determination (which is confused by the possible presence of multiple
populations), on the definition of a homogeneous reddening scale.

Gaia will do its share to improve absolute age by acting on most of
the above items: \\
i) clean CMDs and very accurate photometry at the level 
of the turn-off will allow to obtain a more precise definition and
estimate of the observed V$_{TO}$. Accuracies of $\sim$0.01-0.02 mag
can be foreseen, and are within reach also of the best present and
future observing facilities. \\ ii) The reddening will be monitored by
Gaia for each object as part of the astrophysical parameter
determination and may not be very accurate individually, but the
statistical use of all cluster stars could lead to a rather accurate
mean estimate. To be conservative, we assume a factor two improvement
in the accuracy of the reddening values. \\ iii) Similarly, $[Fe/H]$
will be estimated as part of the astrophysical parameter
determination, and the mean of hundreds/thousands stars will carry
rather low internal errors. iv) The most important contribution,
however, will be on the distance determination, by greatly reducing
the error on distance (e.g. by a factor 10 at 10 kpc). \\ Altogether,
we expect that {\em absolute} ages can be known to $\sim$ 5\% or
better.

\subsubsection{Open Clusters}\label{s:OC} 

Open clusters (OCs) are much looser than GCs and will be completely
observable to the centre by Gaia. Therefore, the same type of analysis
described for GCs in Sections \ref{s:GC} and \ref{s:M3} can be applied
to the entire stellar population for each and all of the presently
known OCs.  In addition, Gaia data may well be able to identify new
(faint/loose) clusters.

With Hipparcos, firm results were obtained only on 7 OCs, whereas the
Pleiades still represent a controversial case.

The Pleiades have been studied extensively in the last decade, and
very similar parallax values have been found by various authors:
$\pi$=7.59$\pm$0.14 mas from MS fitting (\cite[Pinsonneault et
al. 1998]{pins98}), $\pi$=7.69 mas from various methods
(\cite[Kharchenko et al. 2005]{khar05}), $\pi$=7.49$\pm$0.07 mas from
HST-FGS parallaxes of three stars in the inner halo (\cite[Soderblom
et al. 2005]{sod05}).  However, from the new reduction of Hipparcos
data \cite[van Leeuwen (2007)]{fvl07} finds $\pi$=8.18$\pm$0.13 mas.
This discrepancy of about 8\% in the distance determination will be
resolved unambiguously by Gaia. Since all the stars of the Pleiades
are brighter than V=15, they will have {\em individual} parallaxes
determined to better than 0.1-0.2\%, and the distance and internal
stellar distribution will be derived with extremely high precision.

\subsubsection{The distance scale: local calibrators}\label{s:Lcalib}

{\bf $\bullet$ RR LYRAES}

RR Lyrae variable stars are the most traditional standard candles, as
their absolute magnitude can be expressed to a first approximation as
$M_V = \alpha + \beta [Fe/H]$, with $\beta \sim$0.2. However, the
zero-point $\alpha$ of this relation is determined to somewhat lower
accuracy than the slope $\beta$.

Hipparcos measured parallaxes for 126 RR Lyrae stars with $<$V$>$ = 10
to 12.5 mag (750-2500 pc, \cite[Fernley et al. 1998]{fern98}), but
only one star, RR Lyr itself, had a parallax measured to better than
20\%, $\pi$=3.46$\pm$0.64 mas (\cite[van Leeuwen 2007]{fvl07}).
However, the parallax measured by \cite[Benedict et al. (2002)]{ben}
using HST data, $\pi$=3.82$\pm$0.20 mas, leads to a shorter distance
modulus by $\sim$0.2 mag. This discrepancy is far too large and
definitely not acceptable for what is supposed to be the basic
luminosity/distance calibrator and the first step in the cosmic
distance scale.

Gaia will obtain the parallax of RR Lyr to $<$ 0.1\% and the
trigonometric distances to {\em all} the field RR Lyraes within 3 kpc
with {\em individual} accuracy $\sigma(\pi)/\pi <$3\% (better than
30\% for most galactic RR Lyraes).  This will allow to calibrate the
$M_V-[Fe/H]$ relation with very high accuracy, for application to all
stellar systems where a good estimate of the RR Lyrae metallicity and
mean V magnitude is possible. \\

{\bf $\bullet$ CEPHEIDS} 

Cepheids, along with RR Lyrae stars, form the cornerstone of the
extragalactic distance scale.  Classical (Pop I) Cepheids are several
magnitudes brighter than RR Lyraes, and can be observed in many spiral
and irregular galaxies as far as 25 Mpc (thus reaching the Fornax and
Virgo clusters) with the use of the HST and other large ground-based
or space telescopes.  The Hipparcos data provided the first
opportunity to calibrate independently the critical parameters in the
Period-Luminosity-Colour (PLC) relation for classical Cepheids in the
Galaxy.  Hipparcos measured parallaxes for about 250 Cepheids, $\sim$
100 of which with parallax accuracies of 1 mas or less.  With the use
of these data and additional HST parallax measures for 10 of these
stars, \cite[van Leeuwen et al. (2007)]{fvl07_2} derived a new
calibration of the PLC relation leading to a distance modulus for the
LMC of 18.48$\pm$0.03 mag (no metallicity correction), and hence
H$_0$=70$\pm$5 km s$^{-1}$ Mpc$^{-1}$.  This is certainly an excellent
result, but is still affected by uncertainties due to the various
parameters involved in the definition of the calibration itself.

Gaia is expected to measure distances to $<$4\% for all galactic
Cepheids ($<$1\% up to 3 kpc), therefore will provide a definitive
resolution of the controversy about the zero-point of the PLC
relation, as well as about the dependence on period, colour and
metallicity.

Cepheid parallaxes can also be measured by Gaia in extragalactic
systems such as the Sagittarius dwarf galaxy with $\sigma(\pi)/(\pi)
<$ 10\%, and the Magellanic Clouds with $\sigma(\pi)/(\pi)$ $\le$50\%
for all stars with period longer than $\sim$10 days ($M_V \le$ --4.2
mag). \\ This will allow to reach a few fundamental goals: \\
i) define a very accurate PLC relation, including the possible dependence on 
metallicity; \\
ii) establish the distance to the LMC on a completely trigonometric basis, 
and improve its accuracy with the additional help of the Galactic calibration 
relation;  \\ 
iii) establish the universality of the PLC relation, namely its applicability 
to all galaxies and hence the possibility to derive H$_0$ and the age of the 
Universe. \\ 

{\bf $\bullet$ METAL-POOR SUB-DWARFS} 

Within the context of distance determination, metal-poor subdwarfs are
very important as they constitute the reference frame for GC main
sequence stars of similar metallicity. The availability of the high
precision Hipparcos parallaxes prompted numerous determinations of
distances to several Galactic GCs using this Main Sequence Fitting
method.  We refer the reader to e.g. \cite[Gratton et
al. (2003)]{gra03} for a detailed description and review.

However, Hipparcos provided precise enough parallaxes only for $\sim$
30 metal-poor subdwarfs with $M_V \sim$ 5.5 to 7.5 mag (i.e. 2-4 mag
below the TO).  Since the required astrometric accuracy could only be
obtained within limiting magnitude V $\sim$ 10, this allowed to sample
a rather small volume of the local neighborhood within 30-80 pc (and
hence the small number of metal-poor stars).

Gaia's limiting magnitude to V $\sim$ 15 will allow to measure
metal-poor subdwarfs in the same range of absolute magnitude as far as
$\sim$ 800 pc with astrometric accuracy better than $\sim$ 3\%.
Several thousands are expected, thus providing a much better
statistics and finer sampling in metallicity for a more accurate
fitting to any given GC main sequence.

\begin{acknowledgements} This overview of the Gaia project borrows
freely from previous scientific and technical publications, and from
the information available on the Gaia website.  The effort of the many
people involved in the Gaia project is implicit in this synthesis. \\
This work was done under the financial support of ASI grant I/016/07/0 
and PRIN-INAF grant CRA1.06.10.04.   
The author gratefully acknowledges the support of a IAU travel grant. 
\end{acknowledgements}

\end{document}